\documentclass[11pt,letterpaper]{JHEP3}
\usepackage{graphics}
\usepackage{epsfig}
\usepackage{subfigure}
\usepackage{mathrsfs}
\usepackage{amsmath}
\usepackage{amsfonts}
\usepackage{amssymb}
\usepackage{graphicx}
\usepackage{longtable}
\usepackage{rotating}
\usepackage{multirow}

\title{Phase transition  of holographic entanglement entropy  in massive gravity}
\author{Xiao-Xiong Zeng$^{1,2}$, Hongbao Zhang$^{3,4}$, Li-Fang Li$^{5}$\\
$^{1}$School of Material Science and Engineering, Chongqing Jiaotong University, Chongqing,\\
        400074, China\\
$^{2}$Key Laboratory of Frontiers in Theoretical Physics,
       Institute of Theoretical Physics,\\
       Chinese Academy of Sciences,
       Beijing 100190, China\\
$^{3}$  Department of Physics, Beijing Normal University, Beijing, 100875, China\\
$^4$ Theoretische Natuurkunde, Vrije Universiteit Brussel,
and The International Solvay Institutes,
Pleinlaan 2, B-1050 Brussels, Belgium\\
$^{5}$State Key Laboratory of Space Weather, National Space Science Center,\\
       Chinese Academy of Sciences, Beijing 100190, China\\
 \email{xxzeng@itp.ac.cn;  hzhang@vub.ac.be; lilf@itp.ac.cn}
}

 \abstract { The phase structure of holographic entanglement entropy is studied in massive gravity for the quantum systems with finite and infinite volumes, which in the bulk is dual to calculate the minimal surface area for a black hole and black brane respectively. In the entanglement entropy$-$temperature plane, we find for both the black hole and black brane there is a Van der Waals-like phase transition
as the case in thermal entropy$-$temperature plane. That is, there is a first order phase transition for the small charge and  a second order phase transition at the critical
charge. For the first order phase transition, the equal area law is checked and for the second order phase transition, the
critical exponent of the heat capacity is obtained. All the results show that the phase structure of holographic entanglement entropy is the same as that of thermal entropy regardless of the volume of the spacetime  on the boundary.

}

\preprint{}
\begin{document}
\bibliographystyle{ieeetr}
\section{Introduction}

As any ordinary thermodynamic system, black holes have entropy as well as temperature, and obey the thermodynamical laws\cite{Hawking1,Hawking2,Hawking3}. But different from the thermal entropy for the ordinary system, which is proportional to the volume,  the entropy of a black hole is proportional to its area of horizon. Among others, AdS/CFT correspondence offers us a natural explanation for such a seemingly bizarre behavior because it maps a gravitational system to an ordinary system on the boundary with one less dimension\cite{ads1,ads2,ads3}. Moreover, in the context of AdS/CFT, not only can the area of black hole horizon be understood as the thermal entropy for its dual boundary system, but also the area of the bulk co-dimension $2$ minimal surface anchored onto the entangling surface on the boundary  can be used to calculate the corresponding entanglement entropy for the boundary system\cite{Ryu}.

Since the entanglement entropy acquires a similar geometric description to the thermal entropy in the bulk by holography, it is of interest to see whether the entanglement entropy demonstrates the similar behavior as the thermal entropy.  Especially in the recent work \cite{Johnson}, the authors
showed that like the black hole entropy, the entanglement entropy also undergoes a Van der Waals-like phase transition at the same critical temperature  in both the fixed charge ensemble and chemical potential ensemble. They also found the second order phase transition occurring for the entanglement entropy at the same critical point as the thermal entropy with nearly the same value as that from thermal phase transition.

 In light of these interesting results, this work was then generalized to the extended phase space where the cosmological
constant is considered as a thermodynamical variable, and it was found that the entanglement entropy has the similar phase structure as that of the black hole entropy too\cite{Caceres}. Very recently, Nguyen has investigated exclusively the equal area law of holographic entanglement entropy and his result showed that, as black hole entropy, the equal area law also holds  for the entanglement entropy regardless of the size of the entangling surface\cite{Nguyen}. All of these works reinforce the viewpoint that thermal phase transition can also been captured by  the entanglement entropy somehow.

In this paper, we intend to study the phase transition of holographic entanglement entropy in massive gravity. The main motivation of this paper is to explore whether the thermal phase transition in an infinite volume can also be described by holographic entanglement entropy, which has not been reported in the literature until now. Massive gravity, where the graviton is endowed with mass, is one
of the gravity theory that attends to modify general relativity for it is  not UV complete. An important black hole  solution that is ghost-free in massive gravity has been presented recently in \cite{Vegh}. It was found that the graviton
plays the same role as the lattice in the holographic conductor model. In light of this interesting property, Ref.\cite{Cai}
generalized this solution to more general cases and studied the corresponding  thermal phase transition in the grand canonical
ensemble and canonical ensemble.  The $P-V$ criticality was also investigated in the extended phase space\cite{Xu}.
The result shows that the phase transition depends not only on the curvature parameter $k$ but also the graviton mass $m$.
In particular, even for the case $k=0$, the Van der Waals-like phase transition may also occur  provided a proper parameter related to the graviton mass is given.  This interesting character is the main motivation for us to study phase transition of holographic entanglement entropy in this spacetime because it  provides a background with infinite volume on the boundary in  which the Van der Waals-like phase transition can take place.

The organization of this paper is as follows. In the next section, after a brief review of the black hole solution in massive gravity,  we will study the thermal phase transitions and critical phenomena for a black hole and black brane in the fixed charge ensemble,respectively. In both spacetimes, the equal area laws are checked and the critical exponents are calculated. In Section 3, we will concentrate mainly on studying the phase transition of holographic entanglement entropy in the entanglement entropy$-$temperature  plane. For both the black hole and black brane, we find the equal area law is  valid and the critical exponent of the heat capacity in the neighborhood of the critical point is the same as that obtained from the thermal phase transition. The last section is devoted to our conclusions and discussions.


\section{Black hole solutions in massive gravity and thermal phase transition}
The action for an $(n+2)$-dimensional  massive gravity is given by\cite{Vegh}
\begin{equation}
\label{eq1}
S =\frac{1}{16\pi G_{n+2}}\int d^{n+2}x \sqrt{-g} [ R +\frac{n(n+1)}{l^2} -\frac{1}{4}F_{\mu\nu}F^{\mu\nu} +m^2 \sum^4_i c_i {\cal U}_i (g,f)].
\end{equation}
 Here $G_{n+2}$ is the Newton's gravitational constant, which will be set to $1$ later on. In addition, $m$ is the graviton mass, $l$ is the AdS radius,  $f$ is  the reference metric. Here the reference metric is given by $f_{\mu\nu}dx^{\mu}dx^{\nu}=h_{ij}dx^i dx^j$, where $h_{ij}dx^i dx^j$ is the line element for a 2-dimensional Einstein space with constant curvature 2$k$, $k=-1,0,1$ corresponds to
a sphere, flat, or hyperbolic horizon for the spacetime, individually. $F_{\mu\nu}=\partial_{\mu} A_{\nu}-\partial_{\nu} A_{\mu}$,
$c_i$ are constants  and ${\cal U}_i$ are symmetric polynomials of the eigenvalues of the $(n+2)\times (n+2)$ matrix ${\cal K}^{\mu}_{\ \nu} \equiv \sqrt {g^{\mu\alpha}f_{\alpha\nu}}$, i.e.,
\begin{eqnarray}
\label{eq2}
&& {\cal U}_1= [{\cal K}], \nonumber \\
&& {\cal U}_2=  [{\cal K}]^2 -[{\cal K}^2], \nonumber \\
&& {\cal U}_3= [{\cal K}]^3 - 3[{\cal K}][{\cal K}^2]+ 2[{\cal K}^3], \nonumber \\
&& {\cal U}_4= [{\cal K}]^4- 6[{\cal K}^2][{\cal K}]^2 + 8[{\cal K}^3][{\cal K}]+3[{\cal K}^2]^2 -6[{\cal K}^4],
\end{eqnarray}
in which $[{\cal K}]={\cal K}^{\mu}_{\mu}$.  Starting from the action in   (\ref{eq1}), Ref.\cite{Cai} has recently got the solution of black holes in such a  massive gravity as
\begin{equation}
\label{metric}
ds^2 = - f(r) dt^2 + f^{-1}(r) dr^2 + r^2 h_{ij} dx^i dx^j , \ \  i, j =1,\ 2,\ 3, \cdots, n
\end{equation}
  with a chemical potential
   \begin{equation}
 \mu = \frac{Q}{(n-1)r_+^{n-1}},
 \end{equation}
 in which
 \begin{eqnarray}
   \label{solution1}
   f(r) &=& k +\frac{r^2}{l^2} -\frac{M}{r^{n-1}}+\frac{Q^2 }{2n(n-1)r^{2(n-1)} } + \frac{c_0c_1m^2}{n }r+c_0^2c_2m^2  \nonumber \\
       &&   +\frac{(n-1)c_0^3c_3m^2}{r}
 +\frac{(n-1)(n-2)c_0^4c_4m^2}{r^2} \label{metric1},
   \end{eqnarray}
$r_+$ is the event horizon which satisfies $ f(r_+)=0$,  $M$ is the black hole mass, and $Q$ is the black hole charge. In this paper, we only consider  the four$-$dimensional spacetime, namely $c_3=c_4=0$ as shown in \cite{Cai}.
 For convenience, we will relabel the remanent graviton mass dependent parameters,  $c_0c_1m^2/{2}$, $c_0^2c_2m^2$, as $a$, $b$ respectively later in this paper.

 The thermal phase structure of the spacetime in (\ref{metric}) has been investigated extensively\cite{Cai,Xu}. In this paper, we are interested in
 the Van der Waals-like  phase transition\footnote{This behavior can be observed in both the $T-S$ plane in general phase space and  $P-V$ plane in extended phase space, where the cosmological constant is viewed as a
dynamical pressure of the black hole system. It has been pointed out that these two perspectives are related to each other
by a duality similar to the T-duality of string theory\cite{Spallucci}.}. To compare the thermal entropy phase transition with entanglement entropy phase transition directly, we focus only
on the phase transition taking place at the   $T-S$ plane.  In particular we will discuss the phase structure of the thermal entropy and entanglement entropy for both the black hole and black brane, namely  $k=1$ and $k=0$ in (\ref{metric1})\footnote{In principle the case of $k=-1$ can also be analysed, but in practice it is hard to choose a proper entangling surface to implement the calculation.}.

\subsection{The case of $k=1$}
\begin{figure}
\centering
\subfigure[$Q=1.5/3$]{
\includegraphics[scale=0.55]{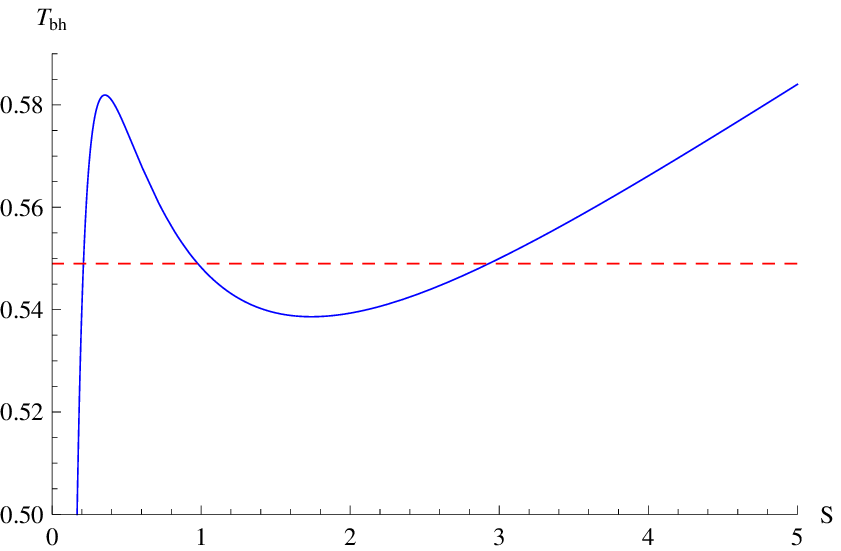}
 }
\subfigure[$Q=2/3$]{
\includegraphics[scale=0.55]{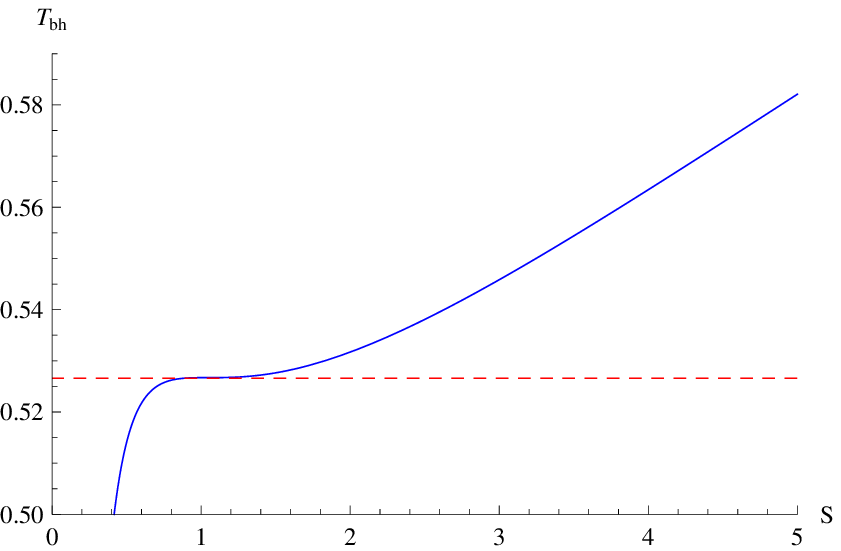}

}
\subfigure[$Q=2.5/3$]{
\includegraphics[scale=0.55]{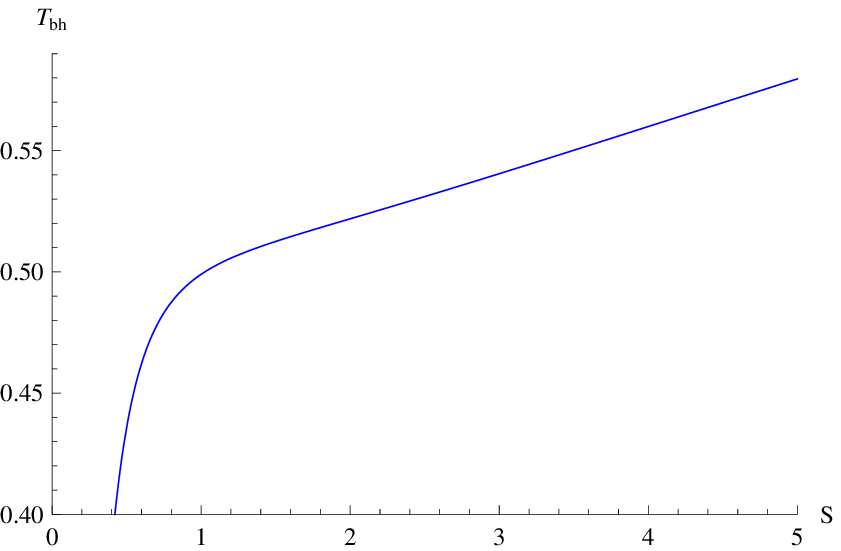}
 }
 \caption{\small Relation between the entropy and temperature for a black hole in a fixed charge ensemble. The red dotted  lines in $ (a)$ and $ (b)$ correspond to the locations of first order phase transition and second order phase transition. } \label{fig1}
\end{figure}

In this case, the  metric  in (\ref{metric1}) can be simplified as
\begin{equation}
f(r)=1-\frac{2 M}{r}+\frac{Q^2}{4 r^2}+\frac{r^2}{l^2}+a r+b.
 \end{equation}
The temperature and entropy of the black hole are
\begin{equation}
T_{bh}=\frac{f(r)^{\prime}}{4\pi}\mid_{r_+}=\frac{12 r_+^4+l^2 \left(-Q^2+4 r_+^2 (1+b+2 a r_+)\right)}{16 l^2 \pi  r_+^3},\label{temperature}
 \end{equation}
\begin{equation}
S=\pi r_+^2.\label{entropy}
 \end{equation}
Substituting (\ref{entropy}) into (\ref{temperature}) and eliminating the parameter $r_+$, we can get the relation between the temperature $T_{bh}$ and entropy $S$, namely
\begin{equation}
T_{bh}=\frac{12 S^2+l^2 \left(-\pi ^2 Q^2+4 (1+b) \pi  S+8 a \sqrt{\pi } S^{3/2}\right)}{16 l^2 \pi ^{3/2} S^{3/2}}.\label{temperatures}
 \end{equation}
Based on this relation, one can plot the isocharge curves on the $T_{bh}-S$ plane for a given charge. The plot is presented in Figure \ref{fig1}, which is obviously similar to that of the Van der Waals phase transition. As can be seen from this plot, for different charges, the curves behave differently. For the small charge in (a), we can see that the curve is not monotonic. According to the definition of the specific heat capacity
\begin{equation}
C_{Q}=T_{bh}\frac{\partial S}{\partial T_{bh}}\mid_Q \label{capacity},
 \end{equation}
we know that the specific heat capacity  changes from positive to negative and then  to positive again. That is, for an intermediate range of event horizon $r_+$ the black hole is thermodynamically unstable. In other words, there are three holes competing thermodynamically. The smallest hole continues to win until a temperature $T_{\star}$, above which the system jumps to the large hole. As stressed in \cite{Chamblin}, this kind of transition is a first order phase transition. When the charge increases to the critical charge, the smallest hole and the largest hole merge into one and squeeze out the unstable black hole. So there is an inflection point in the $T_{bh}-S$ plane, which is shown in (b) of Figure \ref{fig1}.
One can see that the specific heat capacity is divergent at the inflection point and it has been  shown that this phase transition is a second order  phase transition.
\begin{figure}
\centering
\subfigure{
\includegraphics[scale=0.75]{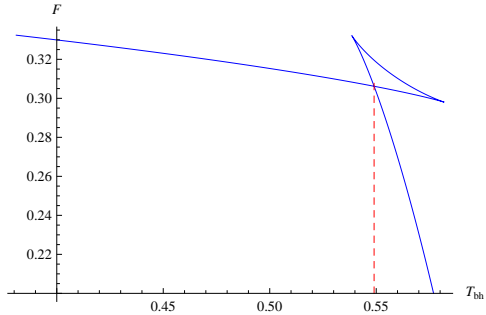} }
 \caption{\small Relation between the free energy and temperature for a black hole at $Q=1.5/3$. The intersection point between the red dotted  line and horizontal coordinate  is the temperature of first order phase transition.} \label{fig2}
\end{figure}
At the inflection  point, the critical charge and critical entropy are determined by the following equations
\begin{equation}
(\frac{\partial T_{bh}}{\partial S})_Q=(\frac{\partial^2 T_{bh}}{\partial S^2})_Q=0. \label{heat}
 \end{equation}
Substituting (\ref{temperatures}) into  (\ref{heat}), we find the critical charge, critical entropy and critical temperature
\begin{equation}
Q_c=\frac{l(1+b)}{3},
 \end{equation}
\begin{equation}
S_c=\frac{l^2(1+b)\pi}{6}, \label{criticalentropy}
 \end{equation}
\begin{equation}
T_c=\frac{4+4 b+\sqrt{6} a \sqrt{(1+b) l^2}}{2 \sqrt{6} \sqrt{(1+b) l^2} \pi } \label{ct}.
 \end{equation}
Obviously these critical parameters  depend on the parameters associated with the graviton mass\footnote{In this paper, we are not interested in the effect of the graviton mass on the critical behavior, for more information about this topic please see  \cite{Cai,Xu}. }. As $a$ and $b$ vanish, these values are the same as that of the AdS-RN black hole. As $b>-1$, we can observe
that there is always a  Van der Waals-like phase transition. In this paper, we choose  $a=b=1$, and for convenience, we also set the AdS radius $l=1$.

For the first order phase transition, we will check whether Maxwell's equal area law holds in this background. The first order transition temperature $T_{\star}$ plays an important role on the  equal area law. To find $T_{\star}$, we first plot the curve about the free energy $F=M-TS$\cite{Cai} with respect to the temperature $T$. The plot  is shown in Figure \ref{fig2}. We see that there is a swallowtail structure, which corresponds to the unstable phase in (a) of Figure (\ref{fig1}). The non-smoothness of the junction indicates that the phase transition is a first order. The critical temperature $T_{\star}$ is apparently given by the horizontal coordinate of the junction. From Figure \ref{fig2}, we find  $T_{\star}=0.549$.
Substituting this temperature into  (\ref{temperatures}), we can get three values of the entropy $S_1=0.210442$,  $S_2=0.977147$, $S_3=2.93091$. With these values, we can now check  Maxwell's equal area law, which states
\begin{equation}
T_{\star}(S_3-S_1)=\int_{S_1}^{S_3}T_{bh}(S,Q)dS. \label{euqalarea}
 \end{equation}
After simplying the whole calculation, we find both the left and right of  (\ref{euqalarea}) equal to 1.49354 exactly. Thus,  Maxwell's equal area law is verified in this background.

For the second order phase transition, we are interested in the critical exponent associated with the heat capacity defined in
 (\ref{capacity}). Near the critical point, we write the entropy as $S=S_c+\delta$, where  $S_c$ is defined in  (\ref{criticalentropy}). Expanding the temperature in small $\delta$, we find
\begin{eqnarray}
T_{bh}=\frac{12S_c^2+l^2 \left(-\pi ^2 Q^2+4 (1+b) \pi  S_c+8 a \sqrt{\pi } S_c^{3/2}\right)}{16 l^2 \pi ^{3/2} S_c^{3/2}}+\frac{\left(12S_c^2+l^2 \pi  \left(3 \pi  Q^2-4 (1+b)S_c\right)\right) \delta}{32 l^2 \pi ^{3/2} S_c^{5/2}}
\nonumber \\
+\frac{3 \left(-4 S_c^2+l^2 \pi  \left(-5 \pi  Q^2+4 (1+b)S_c\right)\right) \delta^2}{128 l^2 \pi ^{3/2} S_c^{7/2}}
+\frac{\left(12 S_c^2+5 l^2 \pi  \left(7 \pi  Q^2-4 (1+b) S_c\right)\right) \delta^3}{256 l^2 \pi ^{3/2} S_c^{9/2}}.\nonumber\\
 \end{eqnarray}
Near the critical point, the first term on the right is the critical temperature, the second and  third terms vanish according to (\ref{heat}). So we get
\begin{eqnarray}
T_{bh}-T_c=\frac{\left(12 S_c^2+5 l^2 \pi  \left(7 \pi  Q^2-4 (1+b) S_c\right)\right) }{256 l^2 \pi ^{3/2} S_c^{9/2}}(S-S_c)^3\label{c1}.
 \end{eqnarray}
With the definition of the heat capacity, we further find  $C_q\sim(T_{bh}-T_c)^{-2/3}$, namely the critical exponent is $-2/3$, which is the same as the one from the mean field theory.

\subsection{The case of $k=0$}
\begin{figure}
\centering
\subfigure[$Q=0.8/3$]{
\includegraphics[scale=0.55]{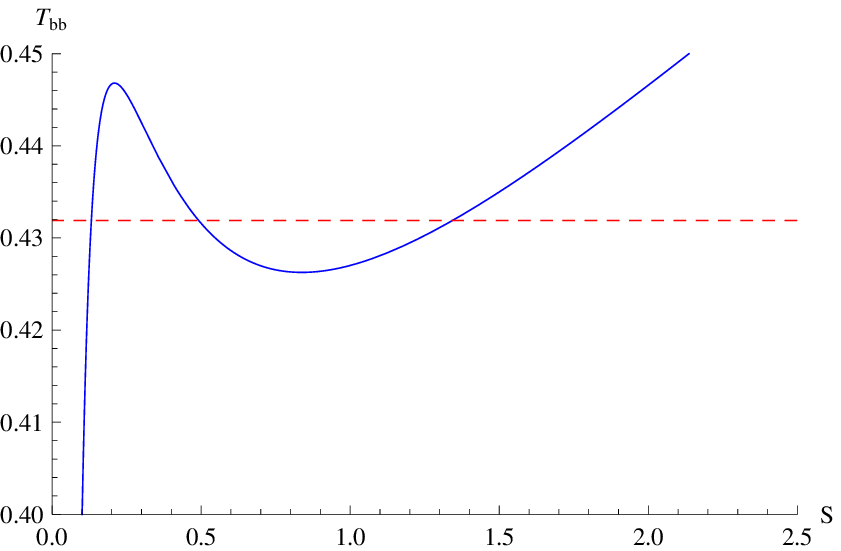}
 }
\subfigure[$Q=1/3$]{
\includegraphics[scale=0.55]{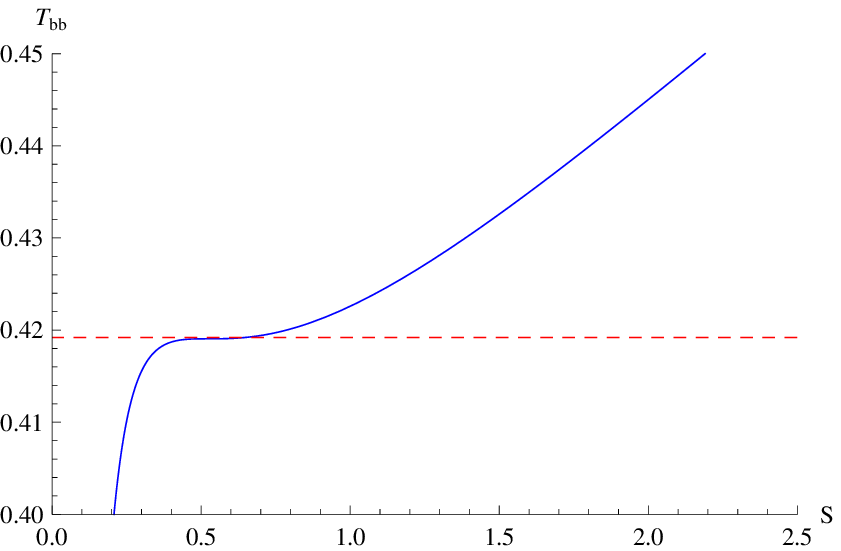}
}
\subfigure[$Q=1.2/3$]{
\includegraphics[scale=0.55]{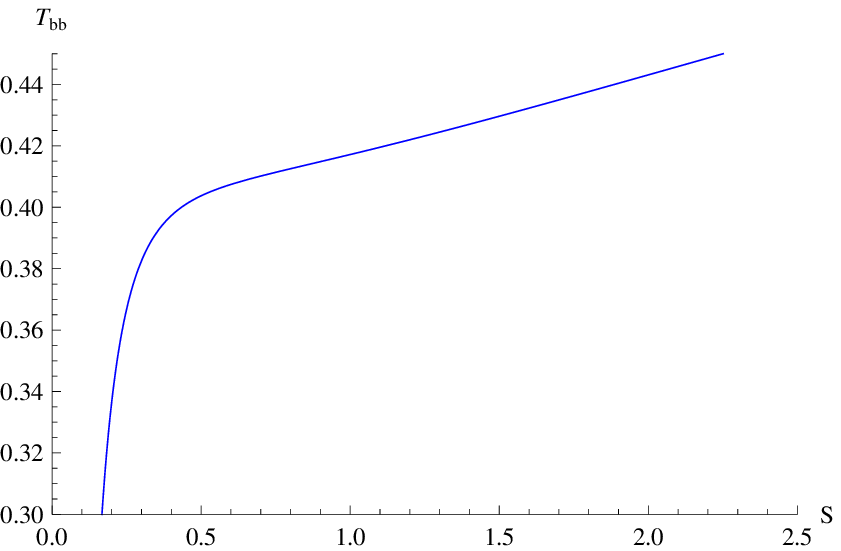}
}
 \caption{\small Relation between the entropy and temperature for a black brane in a fixed charge ensemble. The red dotted  lines in $ (a)$ and $ (b)$ correspond to the locations of first order phase transition and second order phase transition.} \label{fig3}
\end{figure}

For the case $k=0$, the spacetime we are considered is a black brane.
The Hawking temperature is
\begin{equation}
T_{bb}=-\frac{Q^2-4 r_+^2 (b+r_+ (2 a+3 r_+))}{16 \pi  r_+^3}.\label{btemperature}
 \end{equation}
Replacing the event horizon $r_+$ by the  entropy $S$,   (\ref{btemperature}) can be rewritten as
\begin{equation}
T_{bb}=\frac{-\pi ^2 Q^2+4 b \pi  S+8 a \sqrt{\pi } S^{3/2}+12 S^2}{16 \pi ^{3/2} S^{3/2}},\label{cbtemperature}
 \end{equation}
whereby, we can also observe the Van der Waals-like  phase transition, which is presented in Figure \ref{fig3}.
Based on (\ref{heat}), we can also get the critical charge, critical entropy and critical temperature
\begin{equation}
Q_c=b/3,
 \end{equation}
\begin{equation}
S_c=\frac{b \pi }{6}, \label{bcriticalentropy}
 \end{equation}
\begin{equation}
T_c=\frac{3 a+2 \sqrt{6} \sqrt{b}}{6 \pi }.
 \end{equation}
It is obvious that these critical values depend  on the parameters associated with the graviton mass.
\begin{figure}
\centering
\subfigure{
\includegraphics[scale=0.75]{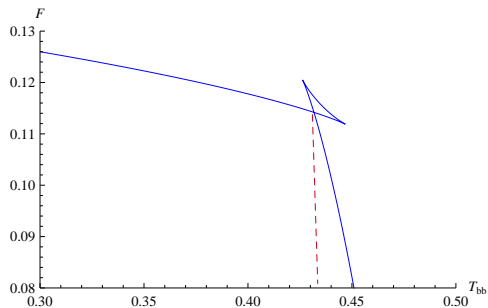} }
 \caption{\small Relation between the free energy and temperature for a black brane at $Q=0.8/3$. The intersection point between the red dotted  line and horizontal ordinate  is the temperature of first order phase transition.} \label{fig4}
\end{figure}
As $a$ and $b$ vanish, the Van der Waals-like phase transition  disappears. That is, in Einstein gravity, this is no Van der Waals-like phase transition  for an AdS-RN black brane.

To check Maxwell's equal area law, we should find the first order phase transition temperature $T_{\star}$ from the $F-T_{bb}$ plot, which is shown in  Figure \ref{fig4}. Reading
from  the horizontal coordinate of the junction, we find  $T_{\star}=0.4316$.
Substituting this temperature into (\ref{cbtemperature}), we get $S_1=0.131235$,  $S_3=1.32697$, where $S_1$,  $S_3$ are the smallest and largest roots of equation $T_{bb}=T_{\star}$. Substituting these values to (\ref{euqalarea}), we find the left equals  0.5161
 and the right equals  0.5162. Obviously both the left and the right are equal nearly within our numerical accuracy, which indicates that Maxwell's equal area law is satisfied in this background.

Similarly we can also get the critical exponent for the second phase transition in (b) of
Figure \ref{fig3}. Writing the entropy as $S=S_c+\delta$ and expanding the temperature in small $\delta$, we find
\begin{eqnarray}
T_{bb}-T_c=\frac{\left(35 \pi ^2 Q^2-20 b \pi  S_c+12 S_c^2\right)}{256 \pi ^{3/2} S_c^{9/2}}(S-S_c)^3,
 \end{eqnarray}
 where we have used (\ref{heat}).
This relation is similar to that of the black hole. That is to say, the critical exponent is $-2/3$ too.

\section{Holographic entanglement entropy and its phase structure}

Having understood the phase structure of a black hole and black brane in massive gravity from the viewpoint of thermodynamics, we will study entanglement entropy in both spacetimes. The main motivation is to check whether entanglement entropy has the similar phase structure as that of the thermal entropy.

For a given quantum field theory described by a density matrix $\rho$, entanglement entropy for a region $A$ and its complement $B$ is defined as
\begin{eqnarray}
S_A=-Tr_A(\rho_A \ln \rho_A),
 \end{eqnarray}
where  $\rho_A=Tr_B(\rho)$ is the reduced density matrix. Usually it is not easy to calculate this quantity in field theory. But from the viewpoint of holography, Ref.\cite{Ryu} gave a very simple geometric description for computing $S_A$ for static states in terms of
the area of a bulk minimal surface anchored on $\partial A$, which states that
\begin{eqnarray}
S_A=\frac{\text{Area}(\gamma)}{4}, \label{ee}
 \end{eqnarray}
where $\gamma$ is the codimension-2 minimal surface with boundary condition $\partial \gamma=\partial A$. Here we will still use (\ref{ee}) to calculate entanglement entropy  and study the corresponding phase structure for preliminary study.

\subsection{Phase transition of Entanglement entropy in a finite volume system}
For a black hole, the space on the boundary is a round. The volume of the space is finite in this case. To avoid the entanglement entropy to be affected by the surface that wraps the horizon, we will choose a small region as $A$. To be more precise,
as done in \cite{Chamblin,Hubeny1}, we choose the region $A$ to be a spherical cap on the boundary delimited by $\theta\leq \theta_0$. In this case the area can be written as
\begin{eqnarray}
A=2 \pi \int_0 ^{\theta_0}\mathcal{L}(r(\theta),\theta) d\theta, \mathcal{L}=r \sin\theta\sqrt{\frac{(r^{\prime})^2}{f(r)}+r^2},
 \end{eqnarray}
in which $r^{\prime}=dr/ d\theta$. Imagining $\theta$ as time, and treating $\mathcal{L}$ as the Lagrangian, one can get the equation of motion for $r(\theta)$ by making use of the Euler-Lagrange equation.
It seems to be impossible to get the analytical solution of  $r(\theta)$, so we will solve it numerically with the boundary conditions $r^{\prime}(0)=0$, $r(0)= r_0$. With the numerical result of  $r(\theta)$, we can easily obtain the area and further entanglement entropy. Note that the area is divergent, so it should be regularized by subtracting off the area in pure AdS with the same entangling surface on the boundary.
\begin{figure}
\centering
\subfigure[$Q=1.5/3$]{
\includegraphics[scale=0.55]{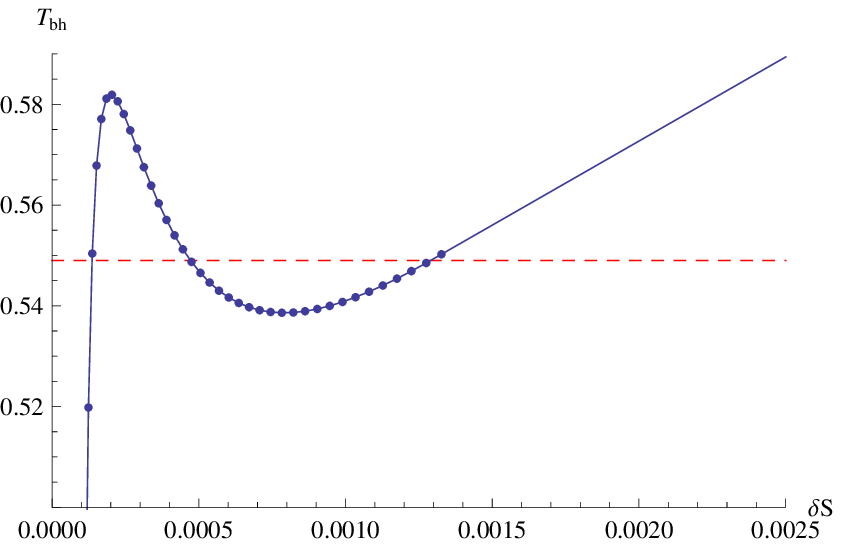}
\label{shell1_RN} }
\subfigure[$Q=2/3$]{
\includegraphics[scale=0.55]{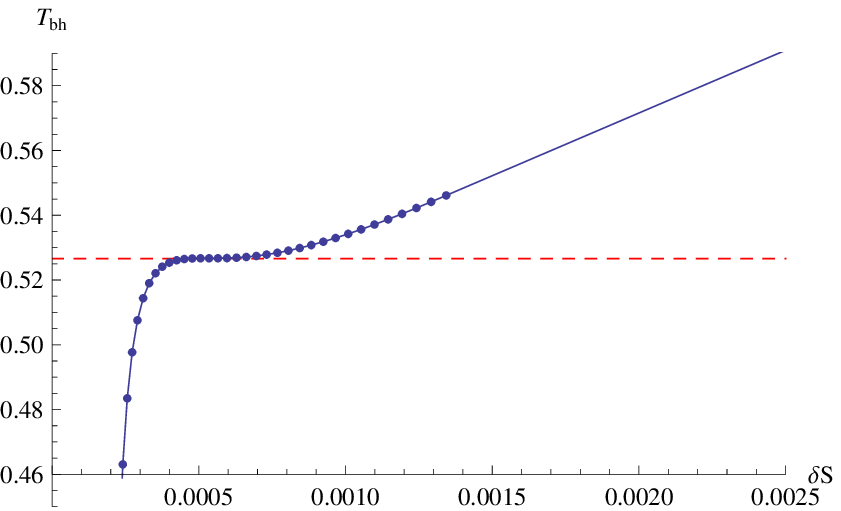}
\label{shell2_RN}
}
\subfigure[$Q=2.5/3$]{
\includegraphics[scale=0.55]{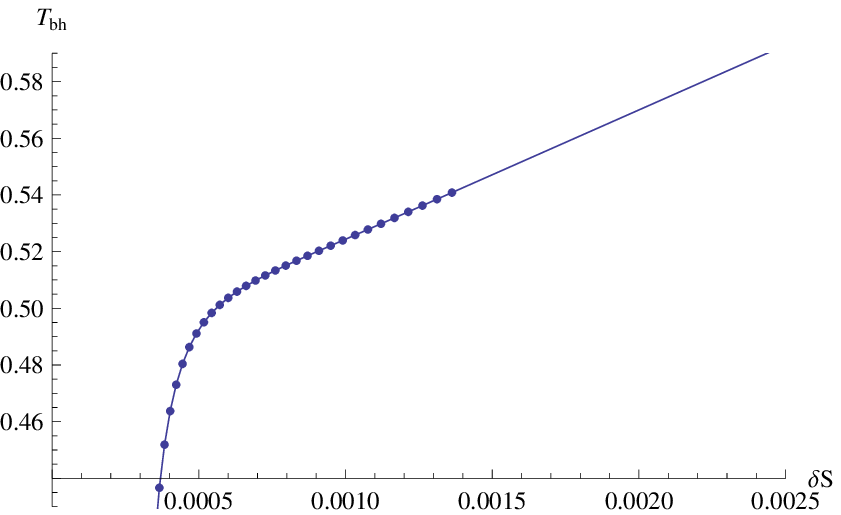}
\label{shell3_RN} }
 \caption{\small Relation between the entanglement entropy and temperature for a black hole in a fixed charge ensemble. The boundary is located at $\theta_0=0.1$. The red dotted  lines in $ (a)$ and $ (b)$ correspond to the locations of first order phase transition and second order phase transition.} \label{fig5}
\end{figure}
\begin{figure}
\centering
\subfigure[$Q=1.5/3$]{
\includegraphics[scale=0.55]{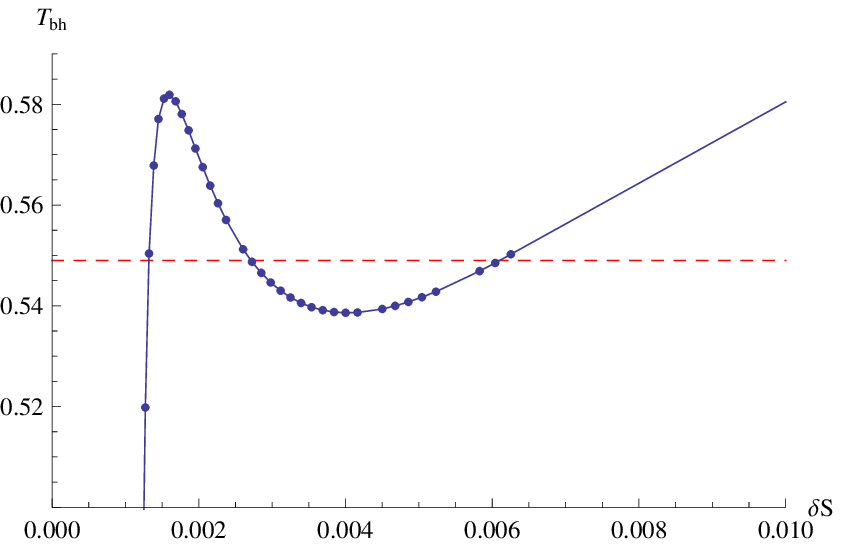}
\label{shell1_RN} }
\subfigure[$Q=2/3$]{
\includegraphics[scale=0.55]{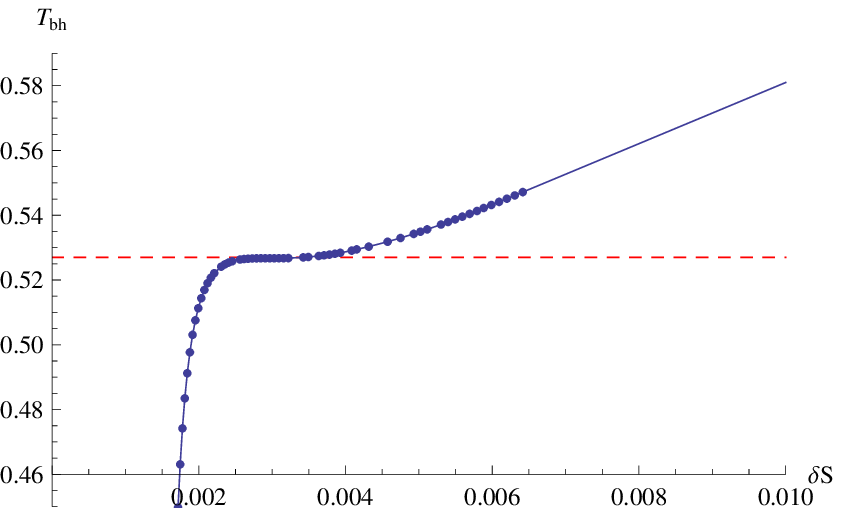}
\label{shell2_RN}
}
\subfigure[$Q=2.5/3$]{
\includegraphics[scale=0.55]{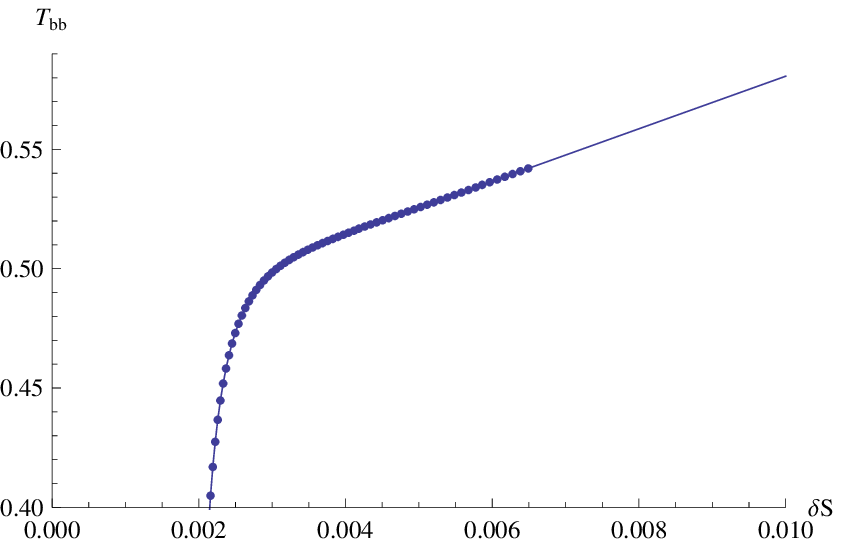}
\label{shell3_RN} }
 \caption{\small Relation between the entanglement entropy and temperature for a black black hole in a fixed charge ensemble.  The boundary is located at $\theta_0=0.16$. The red dotted  lines in $ (a)$ and $ (b)$ correspond to the locations of first order phase transition and second order phase transition.} \label{fig6}
\end{figure}
In Einstein gravity, one can often get an analytical solution of the minimal surface in the pure  AdS. In  our massive gravity, we find that the spacetime for the pure AdS is given by\footnote{Note that in this case, the space is not strictly a pure  AdS, but for convenience we also call it  pure  AdS.}
\begin{eqnarray}
dS^2=-(1+\frac{r^2}{l^2}+a r+b)dt^2+\frac{1}{(1+\frac{r^2}{l^2}+a r+b)}dr^2+r^2(d\theta^2+\sin^2\theta d\phi^2).
 \end{eqnarray}
It is obvious that the parameters related to the graviton mass affect the structure of the pure  AdS. The analytical result for the pure  AdS in Einstein  gravity thus can not be used and  we obtain the corresponding minimal surface for the pure AdS in our massive gravity by numeric strategy too. We label the regularized entanglement entropy as $\delta S$.
For the numerical computation, we choose $\theta_0=0.1$ and $\theta_0=0.16$. The UV cutoff in the dual field theory is set to be $r(0.099)$ and $r(0.159)$ by holography respectively. To compare with the  phase transition of thermal entropy, we will study the relation between the entanglement entropy and Hawking temperature, which is regarded as the temperature of the dual field theory. The numeric results for $\theta_0=0.1$ and $\theta_0=0.16$ are shown in Figure \ref{fig5} and Figure \ref{fig6} respectively. We can see that for a given charge, the relation between the entanglement entropy and temperature is similar to that between the black hole entropy and temperature.
That is, there is a critical charge and below it the first order phase transition occurs. In this case there is an unstable hole interpolating between a small and a large stable hole. As the temperature grows to the critical temperature $T_{\star}$, the small black hole jumps to the large black hole. As the charge grows to the critical charge, the small  hole and the large hole merge so that the unstable hole shrinks into an inflection point where the second order phase transition occurs. For a large enough charge, a large stable black hole forms and the entanglement entropy grows monotonically as the temperature increases.
Comparing Figure \ref{fig5} with  Figure \ref{fig6}, we find the profile  for different sizes of entangling surface seems to be the same, which indicates that the phase structure of entanglement entropy does not depend on the size of entangling surface.

For the first order phase transition of the entanglement entropy, we now check whether  Maxwell's equal area law holds. For the case $\theta_0=0.1$, we first construct an Interpolating Function of the temperature $T(\delta S, Q)$ using the data obtained numerically\footnote{Though the entanglement entropy resembles the black hole entropy, they have different function dependence of temperature.}. At the first order phase transition point, we  find the smallest and largest roots for the equation   $T=T_{\star}$ are $\delta S_1= 0.000136338$, $\delta S_3=0.00128997$. Substituting these values into  the equal area law
\begin{eqnarray}
T_{\star} (\delta S_3-\delta S_1)=\int_{\delta S_1}^{\delta S_3} T(\delta S, Q) d\delta S,\label{arealaw}
 \end{eqnarray}
we find the left equals  0.0006333
 and the right equals  0.0006337. Obviously both the left and the right are equal within our numerical accuracy.  Similarly, for the case $\theta_0=0.16$, we find the left equals 0.002625
 and the right equals  0.002627, which also indicates that the equal area law is satisfied.

Now let's turn to the critical exponent of the second order phase transition of the entanglement entropy. Similar to the definition of specific heat capacity in thermodynamics, we also define an analogous specific  heat capacity for entanglement entropy
\begin{eqnarray}
\mathcal{C}_Q=T_{bh}\frac{\partial \delta S}{\partial T_{bh}}\mid_Q. \label{cheat1}
 \end{eqnarray}
Provided a similar relation as that in (\ref{c1}) is satisfied, then with (\ref{cheat1}) we can get the critical exponent of phase transition of entanglement entropy. Here  we are interested in the logarithm of the quantities $T_{bh}-T_c$, $\delta S-\delta S_c$, in which $T_c$ is the critical temperature in (\ref{ct}) and  $\delta S_c$ is the critical entropy obtained numerically.
\begin{figure}
\centering
\subfigure{
\includegraphics[scale=0.75]{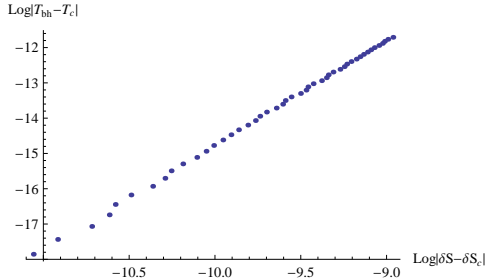}
 }
 \caption{\small Plot of  $\log\mid T_{bh}-T_c\mid$ versus $\log\mid\delta S-\delta S_c\mid $ based on 45 points closest to the
inflection point in Figure 5 $(b)$.} \label{fig9}
\end{figure}
Taking the case $\theta=0.1$ as an example, the linear relation is   shown in Figure  \ref{fig9}  and  the analytical relation can be fitted as
\begin{eqnarray}
\log\mid T_{bh}-T_c\mid=15.0607+2.983 \log\mid\delta S-\delta S_c\mid.
 \end{eqnarray}
Obviously the slope is about 3, therefore the critical exponent for the second order phase transition of entanglement
entropy agrees with that of the thermal entropy.

\subsection{Phase transition of Entanglement entropy in an infinite volume system}

As  have seen in Section 2.2, for a black brane in massive gravity, there exists a Van der Waals-like phase transition, which does't exist in Einstein gravity. So we will study entanglement entropy in this spacetime to check whether the phase structure of entanglement entropy resembles  that of the thermal entropy.

To study the entanglement entropy, we should choose a proper region for $A$.
 Here we choose  a  rectangular  strip parametrized by the boundary coordinates $(x_1,x_2)$ with
the assumption that this infinite  strip is invariant under the $x_2$-direction. In order to make
   the notation as simple as possible, we would like to rename this exceptional coordinate $x_1$ as $x$ and employ it to
    parameterize the minimal area surface such that the entanglement entropy in this background can be written as
\begin{eqnarray}
A=L \int_{-\ell/2} ^{\ell/2} r \sqrt{\frac{(r^{\prime})^2}{f(r)}+r^2} dx,\label{be}
 \end{eqnarray}
in which $L$ is length of the minimal surface along the $x_2$-direction which is set to be 1 here,  $\ell$ is the boundary separation connected  by two boundary points: $(t_0, -\ell/2)$, $(t_0, \ell/2)$, $r^{\prime}=dr/ dx$ and
\begin{equation}
f(r)=-\frac{2 M}{r}+\frac{Q^2}{4 r^2}+\frac{r^2}{l^2}+a r+b.
 \end{equation}
The integrand in
(\ref{be}) can be thought  as the Lagrangian $\cal{L}$ of a fictitious system with $x$ the proper time.
Since the Lagrangian does not depend explicitly on $x$, there is an associated conserved quantity, which
 is useful for us to get the following equation of motion of $r(x)$
\begin{equation}
4r(x)^{2 }f(r)^2-2f(r) r'(x)^2+ r(x) r'(x)^2 f'(r)-2f(r) r r''(x)=0.
 \end{equation}
To solve   $r(x)$, we use the boundary conditions
\begin{equation}
r(0)=r_0, r'(0)=0.
 \end{equation}
\begin{figure}
\centering
\subfigure[$Q=0.8/3$]{
\includegraphics[scale=0.55]{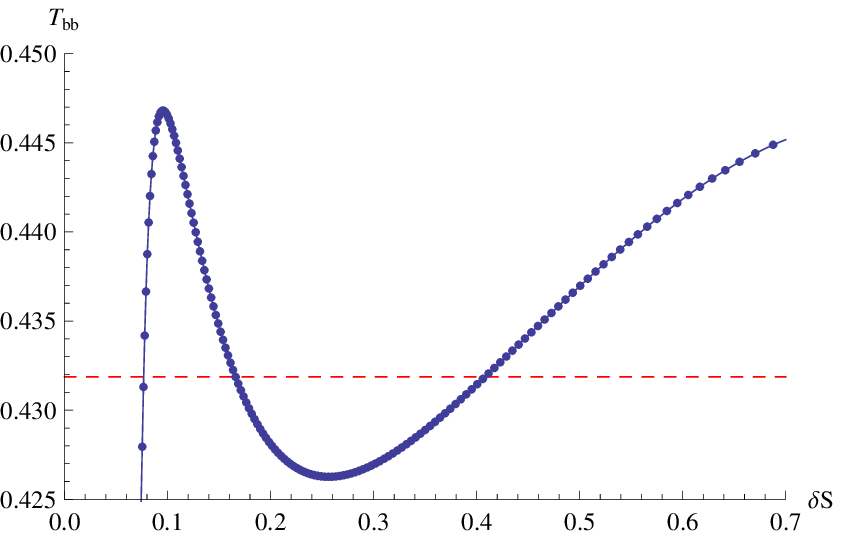}}
\subfigure[$Q=1/3$]{
\includegraphics[scale=0.55]{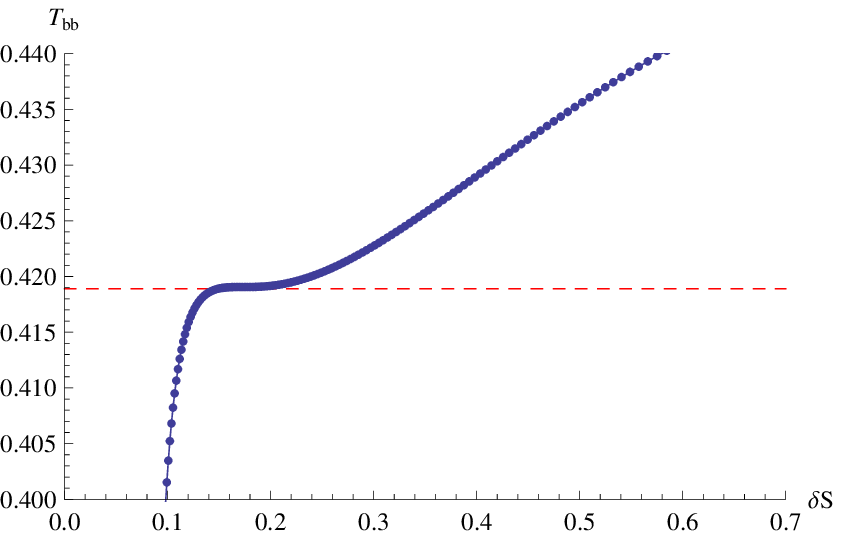}
}
\subfigure[$Q=1.2/3$]{
\includegraphics[scale=0.55]{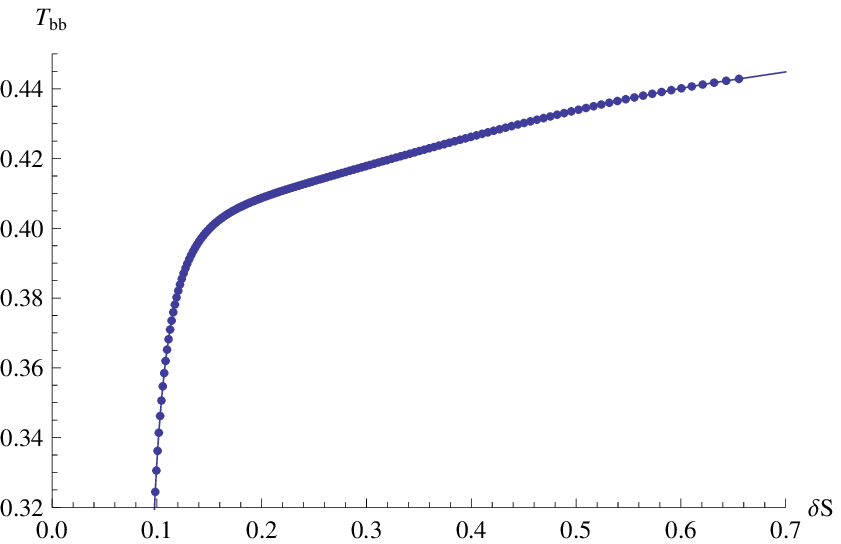}
 }
 \caption{\small Relation between the entanglement entropy and temperature for a black black brane in a fixed charge ensemble.  The boundary separation is $\ell=1.2$. The red dotted  lines in $ (a)$ and $ (b)$ correspond to the locations of first order phase transition and second order phase transition.} \label{fig7}
\end{figure}
\begin{figure}
\centering
\subfigure[$Q=0.8/3$]{
\includegraphics[scale=0.55]{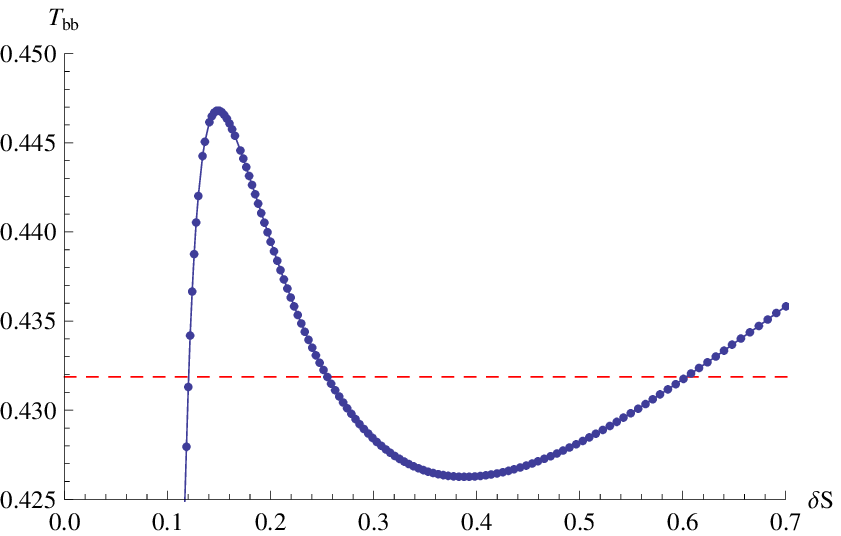}
\label{shell1_RN} }
\subfigure[$Q=1/3$]{
\includegraphics[scale=0.55]{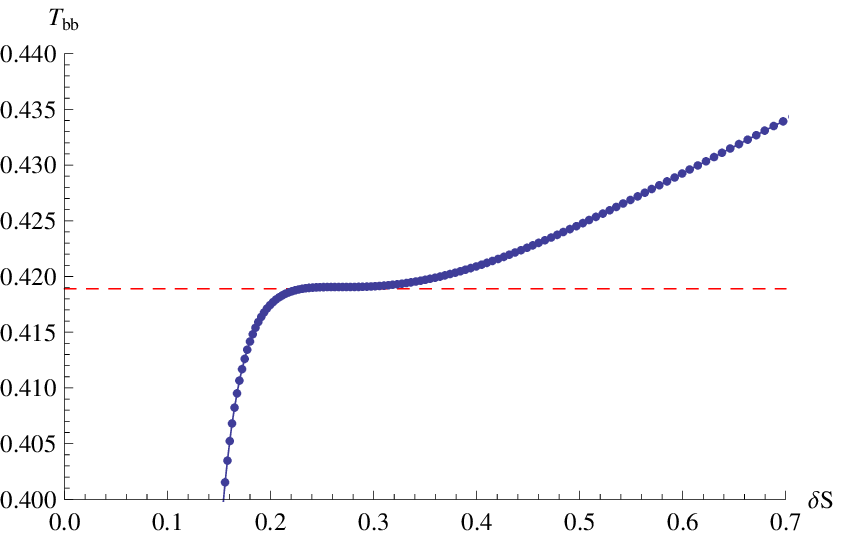}
\label{shell2_RN}
}
\subfigure[$Q=1.2/3$]{
\includegraphics[scale=0.55]{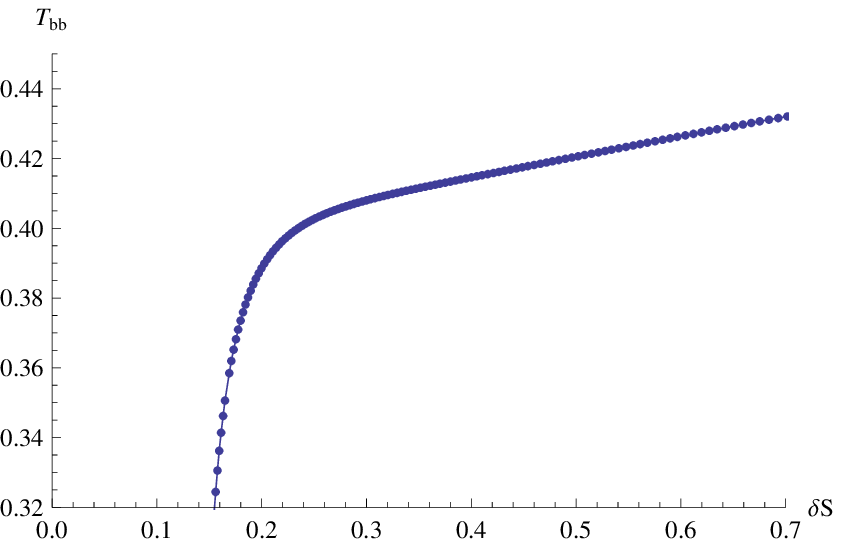}
\label{shell3_RN} }
 \caption{\small Relation between the entanglement entropy and temperature for a black black brane in a fixed charge ensemble.  The boundary separation is $\ell=2.2$. The red dotted  lines in $ (a)$ and $ (b)$ correspond to the locations of first order phase transition and second order phase transition.} \label{fig8}
\end{figure}
Substituting the numerical result of  $r(x)$ into (\ref{be}), we can get the area and further the entanglement entropy. We are also interested in the regularized entanglement entropy $\delta S\equiv S-S_0$, where $S_0$ is entanglement entropy in the pure AdS space of our massive gravity, which is not amenable to analytical calculation but can be obtained by our numerical strategy. In what follows, we are interested in the cases  $\ell=1.2$,  $\ell=2.2$. The UV cutoff is set to be $r(1.19)$ and  $r(2.19)$. The numerical results are shown in  Figure \ref{fig7} and  Figure \ref{fig8}. We can see that the behavior of the curves in Figure \ref{fig7} or  Figure \ref{fig8} is  reminiscent of that in  Figure \ref{fig3}. Namely in the $T_{bb}-\delta S$ phase space, the phase structure of entanglement entropy depends on the charge of the black brane. As the charge grows from small to large enough, the first order phase transition and second order phase transition occur successively, and a stable black brane forms lastly.

We  can also check whether  Maxwell's equal area law holds for the first order phase transition in (a) of  Figure  \ref{fig7} and Figure \ref{fig8}. For the case $\ell=1.2$, the smallest and largest roots for the equation   $T(\delta S, Q)=T_{\star}$  are $\delta S_1= 0.167805$, $\delta S_3=0.403158$, where  $T(\delta S, Q)$ is an Interpolating Function and $T_{\star}$ is the critical temperature of the first order phase transition. Substituting these values into (\ref{arealaw}), we find the left   equals  0.10080
 and the right equals  0.10076.
 Similarly, for the case $\ell=2.2$, the left and right of  (\ref{arealaw}) are found to be $0.20568$ and 0.20565 respectively.
 It is obvious that the equal area law  is satisfied and its validity is independent of the boundary separations.

We  can also get the critical exponent of the heat capacity in the neighborhood of the critical point for the second order phase transition, which is shown in (b) of Figure \ref{fig7} or Figure \ref{fig8}. Taking  the case $\ell=1.2$ as an example. According to the numeric result of the entanglement entropy$-$ temperature curve, we first construct an Interpolating Function  $T(\delta S, Q)$. At the critical temperature, we  obtain the critical entropy $\delta S_c$ by solving the equation   $T(\delta S, Q)=T_{c}$, where $T_{c}$ is defined in
 (\ref{btemperature}).
The relation between  $\log\mid T_{bb}-T_c\mid$ and $\log\mid\delta S-\delta S_c\mid $  is plotted in Figure \ref{fig7}.
\begin{figure}
\centering
\subfigure{
\includegraphics[scale=0.75]{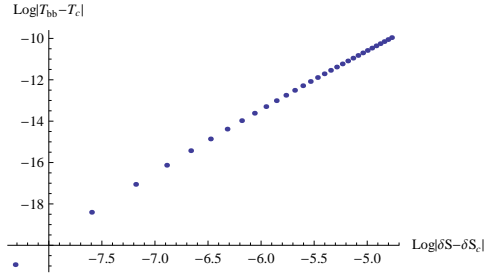}
 }
 \caption{\small Plot of  $\log\mid T_{bh}-T_c\mid$ versus $\log\mid\delta S-\delta S_c\mid $ based on 31 points closest to the
inflection point in Figure 8 $(b)$.} \label{fig10}
\end{figure}
This curve can be fitted as
\begin{eqnarray}
\log\mid T_{bh}-T_c\mid=
6.29083+3.01454\log\mid\delta S-\delta S_c\mid.
 \end{eqnarray}
It is obvious that the slope equals to 3 nearly. In other words, for entanglement
entropy in quantum system with infinite volume, the critical exponent of the heat capacity  also agrees with that of the thermal entropy.

\section{Conclusions}
In this paper, we have studied the phase structure of holographic entanglement entropy in massive gravity. Our result shows that the phase structure of holographic entanglement entropy is the same as that of the thermal entropy regardless  of the volume of a spacetime on the boundary. So one is suggested to use the behavior of holographic entanglement entropy to identify the thermal phase transition. But maybe more importantly, one should be encouraged by these accumulated evidences to find a general proof for such a coincidence between the holographic entanglement entropy and thermal entropy.



\section*{Acknowledgements}
Xiao-Xiong Zeng would like to thank Professor Rong-Gen Cai for his suggestions and encouragement. He is also grateful to Phuc H. Nguyen, Kai Lin, and Shanquan Lan for their  valuable discussions on numerics. Hongbao Zhang is grateful to Minyong Guo for his helpful discussion. He is supported in part by the Belgian Federal Science Policy Office through the Interuniversity Attraction Pole P7/37, by FWO-Vlaanderen through the project G020714N, and by the Vrije Universiteit Brussel through the Strategic Research Program ”High-Energy Physics”. He is also an individual FWO fellow supported by 12G3515N.
This work  is also supported  by the National
 Natural Science Foundation of China (Grant No. 11365008, Grant No. 11205226, Grant No. 11575270), and  Natural Science Foundation of  Education Committee of Chongqing (Grant No. KJ1500530).

\end{document}